\documentclass[12pt]{article}
\usepackage{pdproc,epsfig}
\begin{document}

\renewcommand{\thefootnote}{\alph{footnote}}

\title{Ultra High Energy Cosmic Rays, Z-Shower and Neutrino Astronomy  by Horizontal-Upward Tau Air-Showers\footnote{Invited talk at the
X International Workshop on Neutrino Telescopes, Venice, Italy,
March 11-14, 2003.}}
\author{ Daniele Fargion}

\address{
 Physics department, Universit\'a degli studi "La Sapienza",
Piazzale Aldo Moro 5, 00185 Roma, Italy \\
INFN Roma, Istituto Nazionale di Fisica Nucleare, Italy\\
 {\rm E-mail:daniele.fargion@roma1.infn.it}}

\abstract{ Ultra High Cosmic Rays (UHECR)  Astronomy  may be
correlated to a primary parental Neutrino Astronomy: indeed any
far BL Lac Jet or GRB, sources of UHECR, located  at cosmic
edges, may send its signal, overcoming the severe GZK cut-off, by
help of UHE ZeV energetic neutrino primary. These UHE $\nu$
scattering on relic light ones (spread on wide Hot Local Groups
Halos) maybe fine-tuned $$E_{\nu} =\frac{{M_Z}^2}{2
m_{\overline{\nu}}} \simeq 4\cdot 10^{22} eV \cdot \frac{0.1
eV}{m_{\overline{\nu}}}$$ to combine at once the observed light
neutrino masses and UHECR spectra, leading to a relativistic
Z-Shower in Hot Dark Halos (tens Mpc wide) whose final nuclear
component traces the UHECR event on Earth. Therefore UHECR (with
no longer constrains) may point to far BL Lac sources. This
Z-Burst (Z-Shower) model calls for large neutrino fluxes. Even if
Nature do not follow the present Z-model, UHECR while being
cut-off by Big Bang Radiation, must produce a minimal UHE
neutrino flux, the GZK neutrino secondaries. For both reasons
such UHE Neutrino Astronomy must be tested on Earth. Lowest High
Energy Astronomy is searched by AMANDA, ANTARES underground
detectors by muons tracks. We suggest a complementary higher
energy Neutrino Tau Astronomy inducing Horizontal and Upward Tau
AirShowers. Possible early evidence of such a New Neutrino UPTAUs
(Upward Tau Showers at PeVs energies) Astronomy may be in BATSE
records of Upward Terrestrial Gamma Flashes. Future signals must
be found in  detectors as  EUSO, seeking Upward-Horizontal
events: indeed even minimal, guaranteed, GZK neutrino fluxes may
be observed if EUSO threshold reaches $10^{19}$ eV by enlarging
($3$ meter) its telescope size. }

 \normalsize\baselineskip=15pt

\newpage

-------------------------------------------------

\section{Introduction: UHECR, GZK and Neutrino by EUSO $\nu$}


Ultra High Cosmic Ray, UHECR, may find a solution of their GZK
cut-off by Z-Shower model. Light relic neutrino masses in hot
halos may be the natural calorimeter able to capture far distant
signals in Z-Showers. The Ultra High Energy UHE neutrino
necessary large flux maybe observed on Earth by a new Tau
Astronomy. The same UHE Neutrino may interact on Mountain Chains
or Earth Crust leading to UHE Tau whose decay in flight may be
widely observed. These effects may come as an as-symmetric
Horizontal Showering from the Ande mountain Chain toward AUGER or
by  Horizontal showers flashing upward to new Mountain
Observatory or present and future Gamma Detectors in Space.
Possibly upcoming Tau Showers (UPTAUs) may already blaze the most
sensible GRO Observatory, on Space last decade, leading to first
evidence of Neutrino Astronomy hidden in observed Up-going
Terrestrial Gamma Flashes, by BATSE experiment. Also Horizontal
Tau Showers (HORTAUs), at higher energies may hit the satellite
from the Horizons. Indeed we found possibly early signals in
correlation with EeV galactic anisotropy discovered by AGASA.
Their signals are just at the edge of present AMANDA II
thresholds. Therefore they may already confirm (within a year)
these UPTAUs and HORTAUs fluxes and sources.  There is more and
more expectation also on EUSO outcoming project. This experiment,
while monitoring at dark, downward to the Earth, a wide
atmosphere layers, may discover, among common downward Ultra High
Energy Cosmic Rays, UHECR showers, also first High Energy
Neutrino-Induced Showers. These events are either originated in
Air (EUSO Field of View) or within a widest Earth Crust ring
crown leading to ultra high energy Tau whose decay in flight
produce up-ward and horizontal showers. Upward PeVs neutrinos,
born on air and more probably within a  thin Earth Crust layer,
may shower rarely and blaze to the EUSO detectors. Other higher
energy neutrino may cross horizontally hitting either the Earth
Atmosphere or the Earth Crust:  most of those vertical downward
neutrinos, interacting on air, should be  drown in the dominant
noise of downward UHECR showers. The effective target Masses
originating HORTAUs seen by EUSO may reach (on sea), for most
realistic regime ( at energy $E_{\nu_{\tau}} = 1.2 \cdot 10^{19}$
eV) a huge ring volume $\simeq 2360$ $km^3$. The consequent
HORTAUS event rate (even at $10\%$ EUSO duty cycle lifetime) may
deeply test the expected Z-Burst models by at least a hundred of
yearly events. However, even rarest but inescapable GZK neutrinos
(secondary of photopion production of observed cosmic UHECR)
should be discovered in a dozen of horizontal upward shower
events; in this view an extension of EUSO detectability up to
$\sim E_{\nu}\geq 10^{19}$eV threshold is mandatory. A wider
collecting EUSO telescope (3m diameter) should be considered.
Indeed the very possible discover of an UHECR astronomy, the
solution of the GZK paradox, the very urgent rise of an UHE
neutrino astronomy are among the main goals of EUSO project. This
advanced experiment in a very near future will encompass
AGASA-HIRES and AUGER and observe for highest cosmic ray showers
on Earth Atmosphere recording their tracks from International
Space Station by Telescope facing dawn-ward the Earth. Most of the
scientific community is puzzled by the many mysteries of UHECRs:
their origination because of their apparent isotropy, is probably
extragalactic. Indeed the UHECR spectra injection above $10^19 $eV
seems to be indebted to extra-galactic components. However the
UHECR events are not clustered to any nearby AGN, QSRs or Known
GRBs within the narrow (10-30 Mpc radius) volume defined by the
cosmic 2.75 $K^{o}$ proton drag viscosity (the so called GZK
cut-off \cite{Greisen:1966jv}\cite{Zatsepin:1966jv}). The recent
doublets and triplets clustering found by AGASA seem to favor
compact object (as AGN) over more exotic topological relic
models, mostly fine tuned in mass (GUT, Planck one) and time
decay rate to fit all the observed spectra. However the missing
AGN within a GZK volume is wondering. A possible remarkable
correlation recently shows that most of the UHECR event cluster
point toward BL Lac sources \cite{Gorbunov Tinyakov Tkachev
Troitsky}.

\section{UHECR above GZK  by Z-Burst }
 This
correlation favors a cosmic origination for UHECRs, well above
the near GZK volume. In this frame a relic neutrino mass
\cite{Dolgov2002}, \cite{Raffelt2002}  $m_{\nu} \simeq 0.4$ eV or
($m_{\nu} \simeq 0.1 \div 5$ eV) may solve the GZK paradox
\cite{Fargion Salis 1997} , \cite{Fargion Mele Salis
1999},\cite{Weiler 1999},\cite{Yoshida et all 1998},\cite{Fargion
et all. 2001b},\cite{Fodor Katz Ringwald 2002} overcoming the
proton opacity being ZeV UHE neutrinos transparent (even from
cosmic edges to cosmic photon Black Body drag) while interacting
in resonance with relic neutrinos masses in dark halos (Z-burst or
Z-WW showering models). The light relic neutrinos are the target
(the calorimeter) where UHE $\nu$ ejected by distant (above GZK
cut-off) sources may hit and convert (via Z boson production and
decay) their energy into nuclear UHECR.
\begin{figure}
\centering
\includegraphics[width=.7\textwidth]{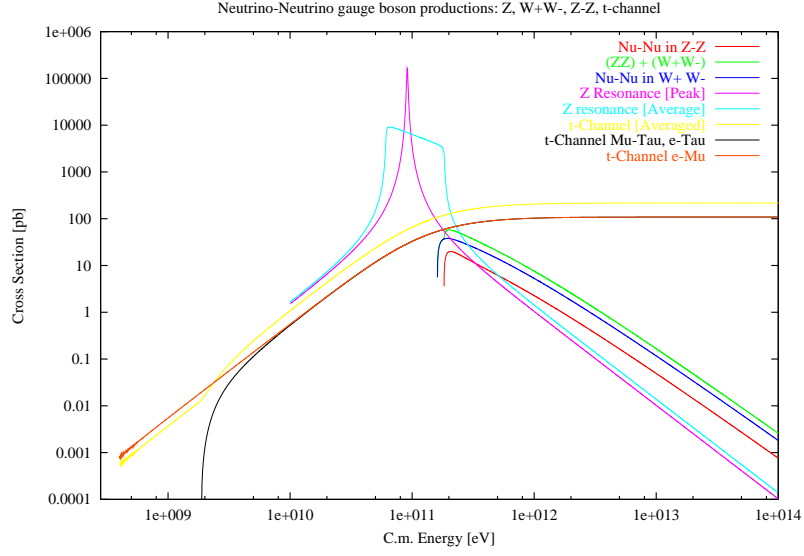}
\caption {The neutrino-relic neutrino cross-sections at center of
mass energy. The Z-peak energy will be smoothed into the
inclined-tower curve, while the WW and ZZ channel will guarantee a
Showering also above a $2 eV$ neutrino masses. The presence of
t-channel plays a role in electromagnetic showering at all
energies above the Z-peak.}\label{fig:fig1}
\end{figure}


\begin{figure}
\centering
\includegraphics[width=.7\textwidth]{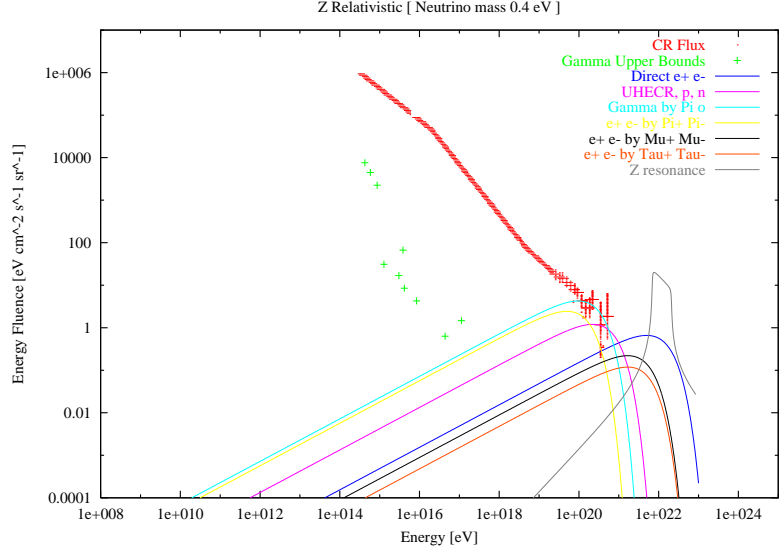}
\caption{Z-Showering Energy Flux distribution for different
    channels assuming a light (fine tuned)  relic neutrino mass
    $m_{\nu} = 0.4 eV$  The total detailed energy percentage
distribution  into neutrino, protons, neutral and charged pions
and consequent gamma, electron pair particles both from hadronic
and leptonic Z, $WW,ZZ$ channels. We also calculated the
electro-magnetic contribution due to the t-channel $\nu_i \nu_j$
interactions as described in previous figure. Most of the $\gamma$
radiation will be degraded around PeV energies by $\gamma \gamma$
pair production with cosmic 2.75 K BBR, or with cosmic radio
background. The electron pairs instead, are  relics of charged
pions losing energies into synchrotron radiation. Note that the
nucleon  injection energy fits the present AGASA data a
corresponding tiny  neutrino mass $m_{\nu}\simeq 0.4 eV$, Lighter
neutrino masses are able to modulate UHECR at higher energies}
\label{fig:fig2}
\end{figure}


\begin{figure}
\centering
\includegraphics[width=.7\textwidth]{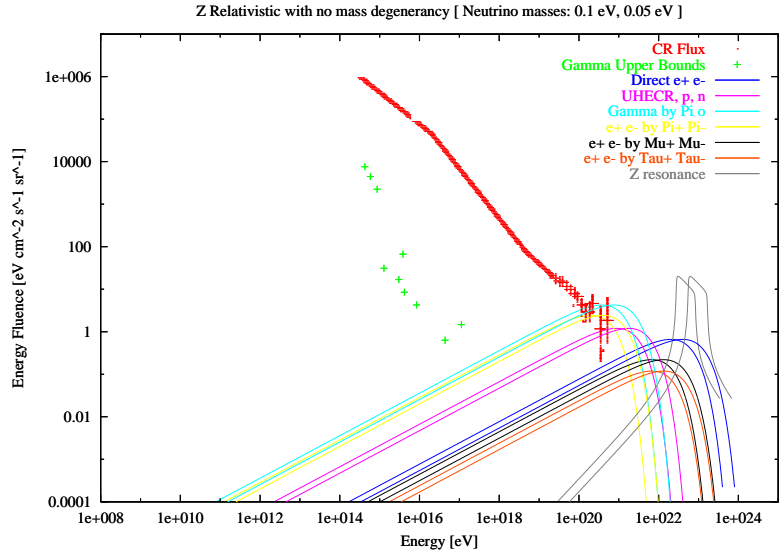}
\caption{Z-Showering Energy Flux distribution for different
channels    assuming a non degenerated twin light  relic neutrino
masses near atmospheric splitting mass values $m_{\nu} = 0.1
eV$,$m_{\nu} = 0.05 eV$ able to partially fill the highest
$10^{20} eV$ cosmic ray   edges.  } \label{fig:fig4}
\end{figure}
\begin{figure}
\centering
\includegraphics[width=.7\textwidth]{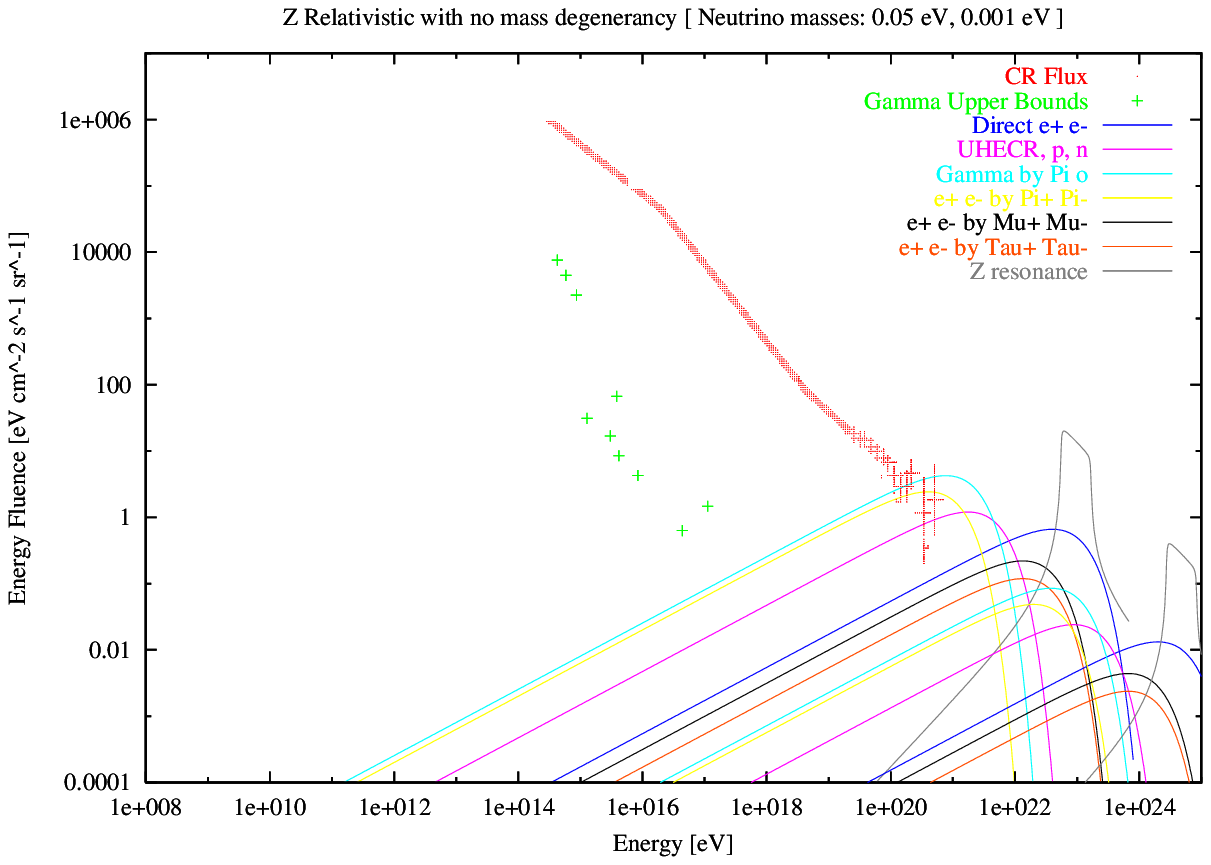}
\caption{Z-Showering Energy Flux distribution for different
channels    assuming a lightest  relic neutrino mass $m_{\nu} =
0.05 eV$ (atmospheric neutrino mass),  $m_{\nu} = 0.001 eV$ ( just
a small fraction (a seventh part) of the minimal solar neutrino
mass split) able to partially fill the highest $10^{20} eV$
cosmic ray  edges. Note that here a realistic suppression for
lightest neutrino density has been assumed. } \label{fig:fig6}
\end{figure}

These light neutrino masses do not solve the galactic or cosmic
dark matter problem but it is well consistent with old and recent
solar neutrino oscillation evidences
\cite{Gallex92},\cite{Fukuda:1998mi},\cite{SNO2002} and most
recent claims by KamLAND \cite{Kamland2002} of anti-neutrino
disappearance  (all in agreement within a Large Mixing Angle
neutrino model and $\triangle {m_{\nu}}^2 \sim 7 \cdot
10^{-5}{eV}^2$) ; these light masses are also in agreement with
atmospheric neutrino mass splitting ($\triangle m_{\nu} \simeq
0.07$ eV) and in fine tune with more recent neutrino double beta
decay experiment mass claim $m_{\nu} \simeq 0.4$ eV
\cite{Klapdor-Kleingrothaus:2002ke}. In this Z-WW Showering for
light neutrino mass models large fluxes of UHE $\nu$ are
necessary,\cite{Fargion Mele Salis 1999},\cite{Yoshida et all
1998}\cite{Fargion et all. 2001b}, \cite{Fodor Katz Ringwald
2002},\cite{Kalashev:2002kx} or higher than usual gray-body
spectra of target relic neutrino or better clustering are needed
\cite{Fargion et all. 2001b}\cite{Singh-Ma}: indeed a heaviest
neutrino mass  $m_{\nu} \simeq 1.2-2.2$ eV while still being
compatible with known bounds (but marginally with more severe WMAP
indirect limits), might better gravitationally cluster leading to
denser dark local-galactic halos and lower neutrino
fluxes\cite{Fargion et all. 2001b}\cite{Singh-Ma}. It should
remarked that in this frame the main processes leading to UHECR
above GZK are mainly the WW-ZZ and the t-channel interactions
\cite{Fargion Mele Salis 1999},\cite{Fargion et all. 2001b}.
These expected UHE neutrino fluxes might and must  be
experienced  in complementary and independent tests.

\newpage

\section{ A new UHE $\nu$ Astronomy by the $\tau$ Showering }

While longest ${\mu}$ tracks in $km^3$ underground detector have
been, in last three decades, the main searched UHE neutrino
signal, Tau Air-showers by UHE neutrinos generated in Mountain
Chains or within Earth skin crust at Pevs up to GZK ($>10^{19}$
eV) energies have been recently proved to be a  new powerful
amplifier in Neutrino Astronomy \cite{Fargion et all 1999}
\cite{Fargion 2000-2002} \cite{Bertou et all 2002} \cite{Hou
Huang 2002} \cite{Feng et al 2002}. This new Neutrino $\tau$
detector will be (at least) complementary to present and future,
lower energy, $\nu$ underground  $km^3$ telescope projects (from
AMANDA,Baikal, ANTARES, NESTOR, NEMO, IceCube). In particular
Horizontal Tau Air shower may be naturally originated by UHE
$\nu_{\tau}$ at GZK energies crossing the thin Earth Crust at the
Horizon showering far and high in the atmosphere \cite{Fargion
2000-2002} \cite{Fargion2001a}  \cite{Fargion2001b} \cite{Bertou
et all 2002} \cite{Feng et al 2002}. UHE $\nu_{\tau}$ are
abundantly produced by flavour oscillation and mixing from muon
(or electron) neutrinos, because of the large galactic and cosmic
distances respect to the neutrino  oscillation ones (for already
known neutrino mass splitting). Therefore EUSO may observe many
of the above behaviours and it may constrains among models and
fluxes and it may also answer open standing questions. We will
briefly enlist, in this  presentation, the main different
signatures and rates of UHECR versus UHE $\nu$ shower observable
by EUSO at 10\% duty cycle time within a 3 year record period,
offering a more accurate estimate of their signals.

\begin{figure}\centering\includegraphics[width=10cm]{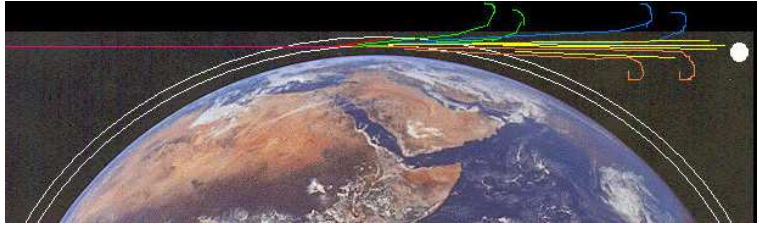}
 \caption {A very schematic Horizontal High
Altitude Shower (HIAS); its fan-like imprint is due to
geo-magnetic bending of charged particles at high quota ($\sim 44
km$). The Shower may  point to a satellite as old gamma GRO-BATSE
detectors or very recent Beppo-Sax,Integral, HETE, Chandra or
future Agile and Swift ones. The HIAS Showers is open and forked
in  five (or three or at least two main component):
($e^+,e^-,\mu^+,\mu^-, \gamma $, or just positive-negative);
these  multi-finger tails may be seen as split tails  by EUSO.}
\label{fig:fig1}
\end{figure}
\begin{figure}
\vspace{- 0.2cm}
\centering\includegraphics[width=10cm]{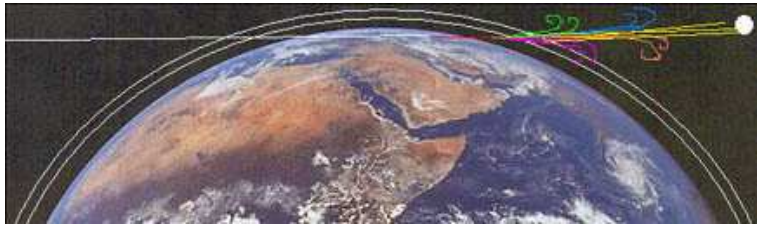}

\caption { As above  Horizontal Upward Tau Air-Shower (HORTAUS)
originated by UHE neutrino skimming the Earth: fan-like jets due
to geo-magnetic bending  shower at high quota ($\sim 23-40 km$):
they may be pointing to an orbital satellite detector . The
Shower tails may be also observable by EUSO just above it.}
\label{fig:fig2bis}
\end{figure}
\begin{figure}
\vspace{-0.2cm}
\centering\includegraphics[width=10cm]{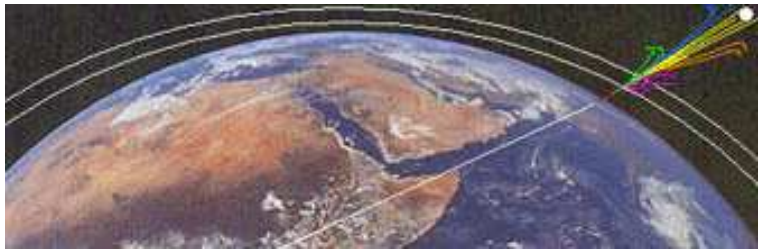}
\caption {A very schematic Upward Tau Air-Shower (UPTAUs)  and its
open fan-like jets due to geo-magnetic bending at high quota
($\sim 20-30 km$). The gamma Shower may be pointing to an orbital
detector. Its
 vertical Shower tail may be spread by
geo-magnetic field into a thin eight-shape beam observable  by
EUSO  as a small blazing oval (few dot-pixels) aligned orthogonal
to the local magnetic field .} \label{fig:fig3}
 \vspace{-0.9cm}
\end{figure}
\begin{figure}
\centering
\includegraphics[width=11cm]{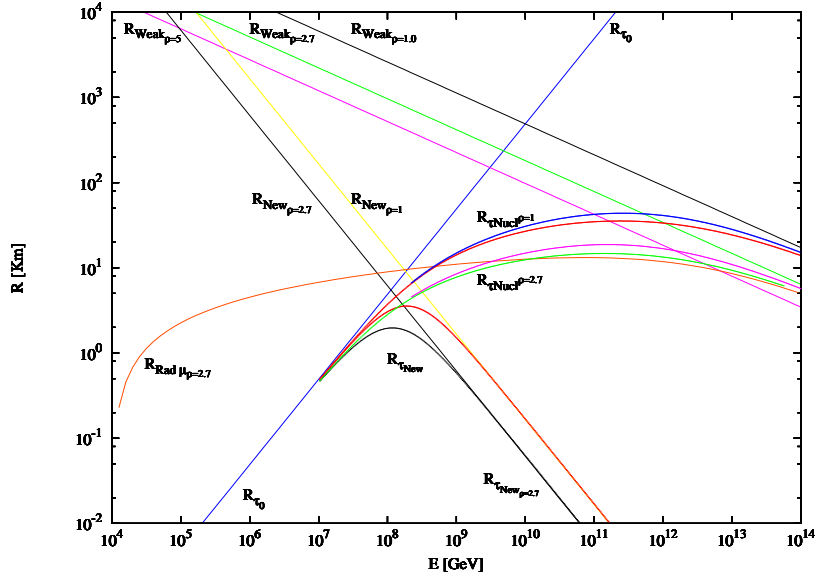}
\caption {Lepton $\tau$ (and $\mu$) Interaction Lengths for
different matter densities: $R_{\tau_{o}}$ is the free $\tau$
length,$R_{\tau_{New}}$ is the New Physics TeV Gravity interaction
range at corresponding densities,$R_{\tau_{Nucl}\cdot{\rho}}$ , is
the combined $\tau$ Ranges keeping care of all known interactions
and lifetime and mainly the photo-nuclear interaction. There are
two slightly different split curves (for each density) by two
comparable approximations in the interaction laws. Note also the
neutrino interaction lenghts above lines $R_{Weak{\rho}}=
L_{\nu}$ due to the electro-weak interactions at corresponding
densities.} \label{fig:fig20}
\end{figure}


\section{Upward UHE $\nu$  Showering in Air versus UPTAUs }
Let us first consider the last kind of Upward $\tau$ signals due
to their interaction in Air or in Earth Crust (UPTAUs). The Earth
opacity will filter mainly  $10^{14}\div{10^{16}}$eV upward events
\cite{Gandhi et al 1998} \cite{Halzen1998} \cite{Becattini Bottai
2001} \cite{Dutta et al.2001} \cite{Fargion 2000-2002}; therefore
only the direct $\nu$ shower in air or the UPTAUs around $3$ PeVs
will be able to flash toward EUSO in a narrow beam ($2.5 \cdot
10^{-5}$ solid angle) jet blazing apparently at
$10^{19}\div{10^{20}}$eV energy. The shower will be opened in a
fan like shape and it will emerge from the Earth atmosphere
spread as a triplet or multi-dot signal aligned orthogonal to
local terrestrial magnetic field. This rare multi-dot
$polarization$ of the outcoming shower will define a
characteristic  signature easily to be revealed. However the
effective observed air mass by EUSO is not $\ 10\%$ (because duty
cycle) of the inspected air volume $\sim 150 km^3$, but because
of the narrow blazing shower cone it corresponds to only to
$3.72\cdot 10^{-3}$ $km^3$. The target volume  increases for
upward neutrino Tau interacting vertically in Earth Crust in last
matter layer (either rock or water), making upward relativistic
$\simeq 3 PeVs$ $\tau$ whose decay in air born finally an UPTAUs;
in this case the effective target mass is (for water or rock)
respectively $5.5\cdot 10^{-2}$$km^3$ or $1.5 \cdot10^{-1}$
$km^3$.


\begin{figure}\centering\includegraphics[width=12cm]{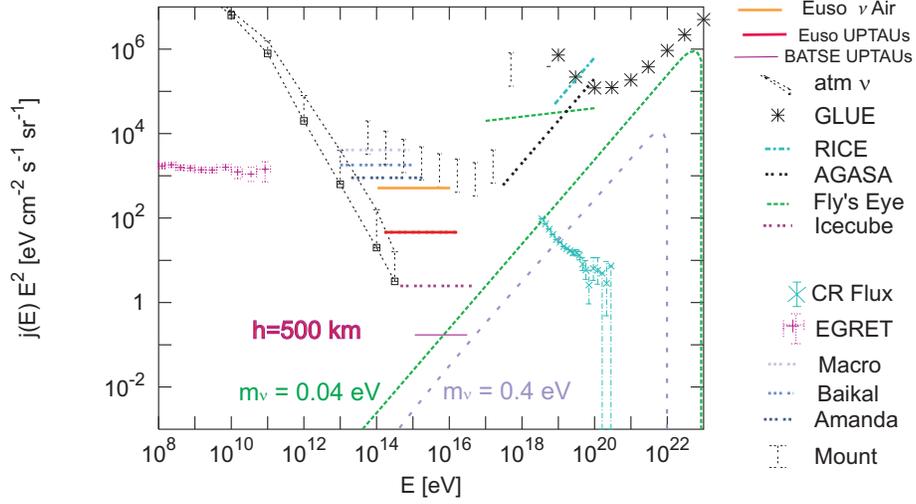}
\caption {Upward Neutrino Air-Shower and Upward Tau Air-shower,
UPTAUs, Gamma and Cosmic Rays Fluence Thresholds and bounds in
different energy windows for different past and future detectors.
The UPTAUs threshold for EUSO has been estimated for a three year
experiment lifetime. BATSE recording limit is also shown from
height $h = 500km$ and for ten year record. Competitive experiment
are also shown as well as the Z-Shower expected spectra in light
neutrino mass values $m_{\nu} = 0.4, 0.04$ eV. }
\end{figure}



The characteristic neutrino interaction are partially summirized
in  Fig.$8$. The consequent $\tau$ and $\mu$ interactions lenght
are also displayed. These volume are not extreme. The consequent
foreseen thresholds for UPTAUs are summirized for $3$ EUSO years
of data recording in Fig.$9$. The UPTAUs signal is nearly $15$
times larger than the Air-Induced Upward  $\nu$ Shower hitting
Air. A more detailed analysis may show an additional factor three
(due to the neutrino flavours) in favor of Air-Induced Showers,
but the more transparent role of PeV multi-generating upward
$\nu_{\tau}$ while crossing the Earth, makes the result
summirized in Fig.$9$. The much wider acceptance of BATSE respect
EUSO and the consequent better threshold (in BATSE) is due to the
wider angle view of the gamma detector, the absence of any
suppression factor as in EUSO duty cycle, as well as the $10$
(for BATSE) over $3$ (for EUSO) years assumed of record
life-time. Any minimal neutrino  fluence $\Phi_{\nu_{\tau}}$ of
PeVs energetic neutrino:
 $$ \Phi_{\nu_{\tau}}\geq 10^2 eV cm^{-2} s^{-1}$$ might be detectable by EUSO.




\begin{figure}\centering\includegraphics[width=12cm]{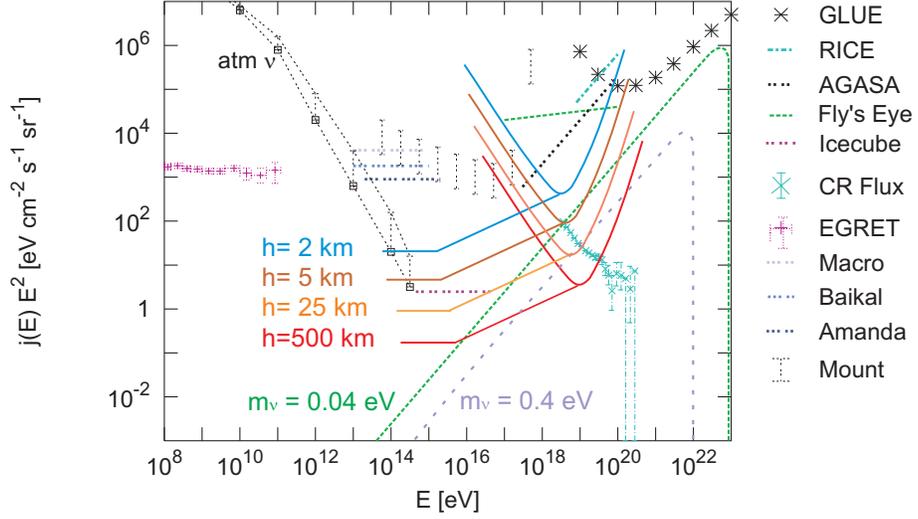}
\caption {UPTAUs (lower bound on the center) and HORTAUs (right
parabolic  curves)  sensibility at different observer heights h
($2,5,25,500 km $) looking at horizons toward Earth seeking upward
Tau Air-Showers adapted over a present neutrino flux estimate in
Z-Shower model scenario for light ($0.4-0.04$ eV) neutrino masses
$m_{\nu}$; two corresponding density contrast for relic light
neutrino masses has been assumed; the lower parabolic bound
thresholds are at different operation height, in Horizontal
(Crown) Detector facing toward most distant horizons edge; these
limits are fine tuned (as discussed in the text) in order to
receive Tau in flight and its
 Shower in the vicinity of the detector; we are assuming a
duration of data records of a decade comparable to the BATSE
record data . The parabolic bounds on the EeV energy range in the
right sides are nearly un-screened by the Earth opacity while the
corresponding UPTAUs bounds  in the center below suffer both of
Earth opacity as well as of a consequent shorter Tau interaction
lenght in Earth Crust, that has been taken into account. }
\label{fig:fig24}
\end{figure}


\newpage

\section{Downward and Horizontal UHECRs}

Let us now briefly reconsider the nature of common Ultra High
Cosmic Rays (UHECR) showers. Their  rate  will offer a useful test
for any additional UHE neutrino signals. Let us assume for sake
of simplicity a characteristic opening angle of EUSO telescope of
$30^o$ and a nominal satellite  height of $400$ km, leading to an
approximate atmosphere area under inspection of EUSO $\sim 1.5
\cdot 10^5 km^2$. Let us discuss the UHECR shower: It has been
estimated (and it is easy to verify)  a $\sim 2\cdot10^{3}$
event/year rate above $3\cdot10^{19}$ eV. Among these "GZK" UHECR
(either proton, nuclei or $\gamma$) nearly $7.45\%\approx 150$
event/year will shower in Air Horizontally with no Cherenkov hit
on the ground. The critical angle $6.7^o$ corresponding to
$7.45\%$ of all the events, is derived from first interacting
quota (here assumed for Horizontal Hadronic Shower near $44$ km
following \cite{Fargion 2000-2002}  \cite{Fargion2001a}
\cite{Fargion2001b}): Indeed the corresponding horizontal edge
critical angle $\theta_{h}$ $=$ $6.7^o$ below the horizons
($\pi{/2}$) is found (for an interacting height h near $44$ km):

\begin{center}
$ {\theta_{h} }={\arccos {\frac {R_{\oplus}}{( R_{\oplus} +
h_1)}}}\simeq 1^o \sqrt{\frac {h_{1}}{km}} $
\end{center}


\begin{figure}\centering\includegraphics[width=12cm]{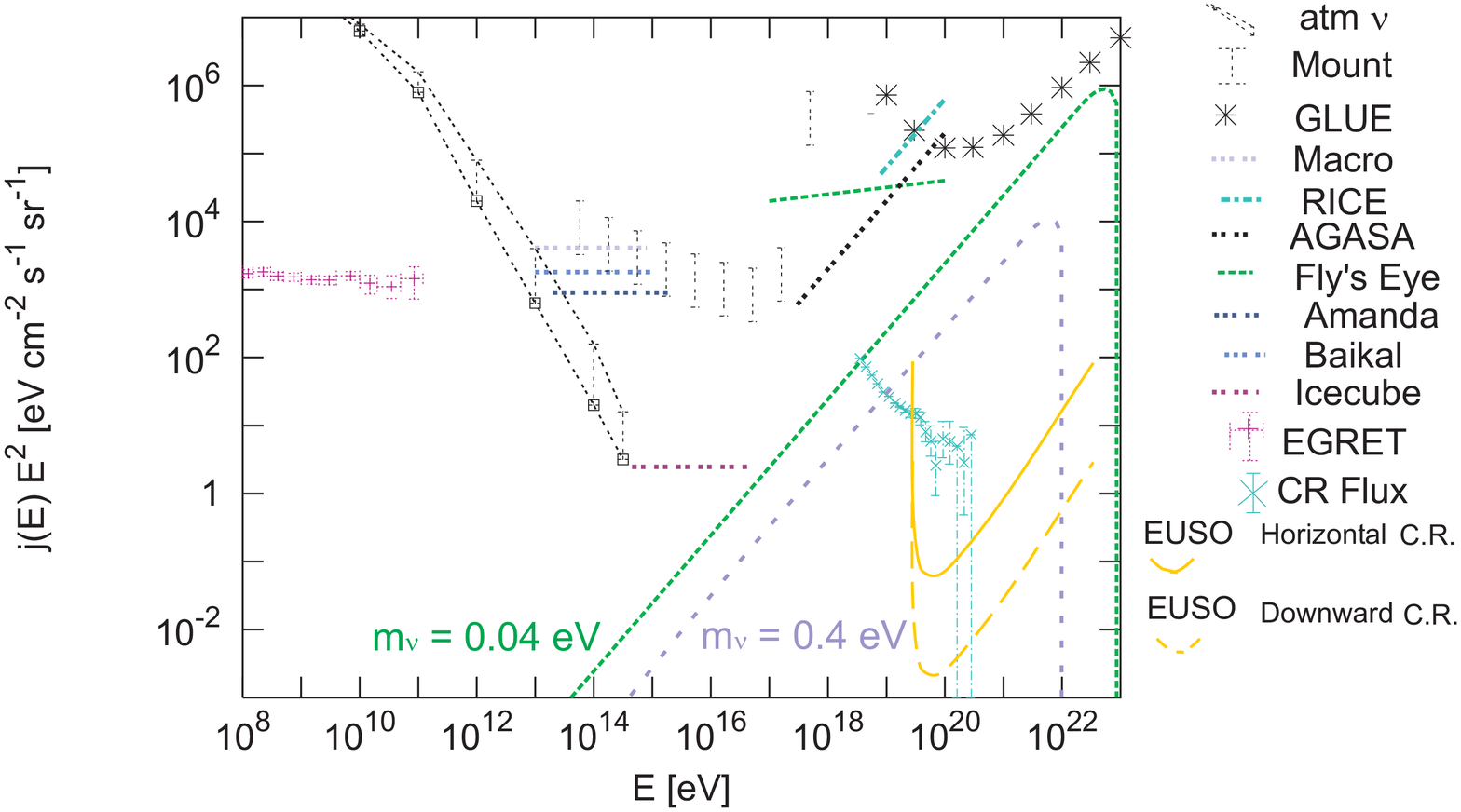}
\caption {Neutrino, Gamma and Cosmic Rays Fluence Thresholds and
bounds in different  energy windows. The Cosmic Rays Fluence
threshold for EUSO has been estimated  for a three year
experiment lifetime. The paraboloid bound shape threshold may
differ upon the EUSO optics and acceptance. Competitive
experiment are also shown as well as the Z-Shower expected
spectra in light mass values.} \label{fig:fig5}
\end{figure}

These Horizontal High Altitude Showers (HIAS) \cite{Fargion2001a}
\cite{Fargion2001b}, will be able to define a new peculiar
showering, mostly very long (hundred kms) and bent and forked (by
few or several degrees) by local geo-magnetic fields. The total
UHECR above $3\cdot10^{19}$ eV will be $\sim 6000$ UHECR and
$\sim 450$ Horizontal Shower within 3 years; these latter
horizontal signals are relevant because they may mimic
Horizontal  induced $\nu$ Air-Shower, but mainly at high quota
($\geq 30-40 km$) and down-ward. On the contrary UHE neutrino tau
showering, HORTAUs, to be discussed later, are also at high quota
($\geq 23 km$), but  upward-horizontal. Their outcoming angle
will be ($\geq 0.2^o-3^o$) upward. Therefore a good angular
($\leq 0.2-0.1 ^o$) resolution to distinguish between the two
signal will be a key discriminator. While Horizontal UHECR are an
important piece of evidence in the UHECR calibration and its GZK
study , at the same time they are a severe back-ground noise
competitive with Horizontal-Vertical GZK Neutrino Showers
originated in Air, to be discussed below. However
Horizontal-downward UHECR are not confused with upward Horizontal
HORTAUs by UHE neutrinos to be summirized in last section. Note
that Air-Induced Horizontal UHE neutrino as well as all down-ward
Air-Induced UHE $\nu$ will shower mainly at lower altitudes
($\leq 10 km$) ; however they are respectively only a small
($\leq 2\% $, $\leq 8\%$) fraction than HORTAUs showers to be
discussed in the following. An additional factor $3$ due to their
three flavour over $\tau$ unique one may lead to respectively
($\leq 6\% $, $\leq 24\% $) of all HORTAUs events: a contribute
ratio that may be in principle an useful test to study the
balanced neutrino flavour mixing.

\section{Air Induced UHE $\nu$ Shower}

UHE $\nu$ may hit an air nuclei and shower vertically or
horizontally or more rarely nearly up-ward: its trace maybe
observable by EUSO preferentially in inclined or horizontal
case.  Indeed  Vertical Down-ward  ($\theta \leq 60^o$) neutrino
induced Air Shower  occur mainly at lowest quota and they will
only partially shower their UHE $\nu$ energy because of the small
slant depth ($\leq 10^3 g cm^{-2}$) in most vertical down-ward
UHE $\nu$ shower. Therefore the observed  EUSO air mass ($1500
km^3$, corresponding to a $\sim 150$ $km^3$ for $10\%$ EUSO
record time) is only ideally the UHE neutrino calorimeter.
Indeed  inclined $\sim{\theta\geq 60^o }$) and horizontal
Air-Showers ($\sim{\theta\geq 83^o }$) (induced by GZK UHE
neutrino) may reach their maximum output  and their event maybe
observed ; therefore only a small fraction ($\sim 30\%$
corresponding to $\sim 50$ $km^3$ mass-water volume for EUSO
observation) of vertical downward UHE neutrino may be seen by
EUSO. This signal may be somehow hidden or masked by the more
common down-ward UHECR showers.  The key reading  signature will
be the shower height origination: $(\geq 40 km)$ for most
downward-horizontal UHECR,$(\leq 10 km)$ for most
inclined-horizontal Air UHE $\nu$ Induced Shower. A corresponding
smaller fraction ($\sim 7.45\%$) of totally Horizontal UHE
neutrino Air shower, orphan of their final Cherenkov flash, in
competition with the horizontal UHECR, may be also clearly
observed: their observable mass is only $V_{Air-\nu-Hor}$ $\sim
11.1$ $km^3$ for EUSO observation duty-cycle.  A more  rare, but
spectacular, double $\nu_{\tau}$-$\tau$ bang in Air (comparable
in principle to the PeVs expected  "double bang" in water
\cite{Learned Pakvasa 1995}) may be exciting, but very
difficult to be observed. \\



The EUSO effective calorimeter mass for such Horizontal event is
only $10\%$ of the UHE $\nu$ Horizontal ones (($\sim 1.1$ $
km^3$)); therefore its event rate is already almost excluded
needing a too high neutrino fluxes (see \cite{Fargion 2002e});
indeed it should be also noted that the EUSO energy threshold
($\geq 3\cdot 10^{19}$eV) imply such a very large ${\tau}$
Lorents boost distance; such large ${\tau}$ track exceed (by more
than a factor three) the EUSO disk Area diameter ($\sim 450$km);
therefore the expected Double Bang Air-Horizontal-Induced ${\nu}$
Shower thresholds are suppressed by a corresponding factor. More
abundant single event Air-Induced ${\nu}$  Shower (Vertical or
Horizontal)  are facing different Air volumes and  quite
different visibility. It must be taken into account an additional
factor three (for the event rate) (because of three light
neutrino states) in the Air-Induced ${\nu}$  Shower arrival flux
respect to incoming $\nu_{\tau}$ (and $\bar{\nu_{\tau}}$ ) in
UPTAUs and HORTAUs, making the Air target not totally a
negligible calorimeter. The role of air Air-Shower will be
discussed elsewhere.

\subsection{ UHE $\nu_{\tau}-\tau$  Air  Single Bang Shower }


\begin{figure}\centering\includegraphics[width=12cm]{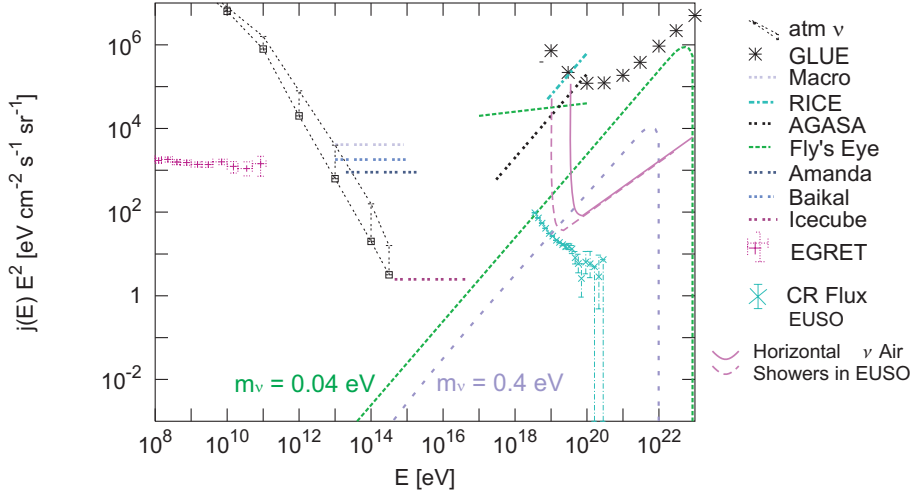}
\caption {EUSO thresholds for Horizontal and Vertical Downward
Neutrino Air induced shower over other $\gamma$, $\nu$ and Cosmic
Rays (C.R.) Fluence  and bounds. The  Fluence threshold for EUSO
has been estimated for a three year experiment lifetime.
Competitive experiments are also shown as well as the Z-Shower
expected spectra in light neutrino mass values ($m_{\nu} = 0.04,
0.4$ eV). } \label{fig:fig7}
\end{figure}


There are also a sub-category  of $\nu_{\tau}$ - $\tau$ "double
 bang" due to a first horizontal UHE $\nu_{\tau}$ charged current interaction
 in air  nuclei (the first bang) that is lost from the EUSO view;
 their UHE  secondary $\tau$ fly and decay leading to a Second Air-Induced Horizontal Shower, within the EUSO
 disk area. These  horizontal "Double-Single $\tau$ Air Bang"  Showers
 (or if you like popular terminology, these Air-Earth Skimming neutrinos or just Air-HORTAU event)
  are produced within a very wide Terrestrial Crown Air Area whose radius is exceeding $\sim 600- 800$ km
 surrounding  the EUSO Area of view. However it is easy to show
 that they will just double the  Air-Induced ${\nu}$  Horizontal Shower
 rate due to one unique flavour. Therefore the total Air-Induced Horizontal Shower (for
 all $3$ flavours and the additional $\tau$ decay in flight) are summirized and considered
 in next figure. The relevant UHE neutrino signal, as discussed below, are due to the
 Horizontal Tau Air-Showers originated within the (much denser)Earth
 Crust:the called  HORTAUs (or Earth Skimming $\nu_{\tau}$).\\
\newpage

\section{UHE $\nu_{\tau}-\tau$ from Earth Skin: HORTAUs}

As already mention the UHE $\nu$ astronomy maybe greatly
amplified by $\nu_{\tau}$ appearance via flavour mixing and
oscillations. The consequent scattering of $\nu_{\tau}$ on the
Mountains or into the Earth Crust may lead to Horizontal Tau
Air-Showers :HORTAUs (or so called Earth Skimming Showers
\cite{Fargion2001a} \cite{Fargion2001b} \cite{Fargion 2000-2002}
\cite{Feng et al 2002}). Indeed UHE $\nu_{\tau}$ may skip below
the Earth and escape as $\tau$ and finally decay in flight,
within air atmosphere, as well as inside  the Area of  view of
EUSO, as shown in Fig. below.

\subsection{UPTAUs and HORTAUs effective Volumes}

\begin{figure}\centering\includegraphics[width=8cm]{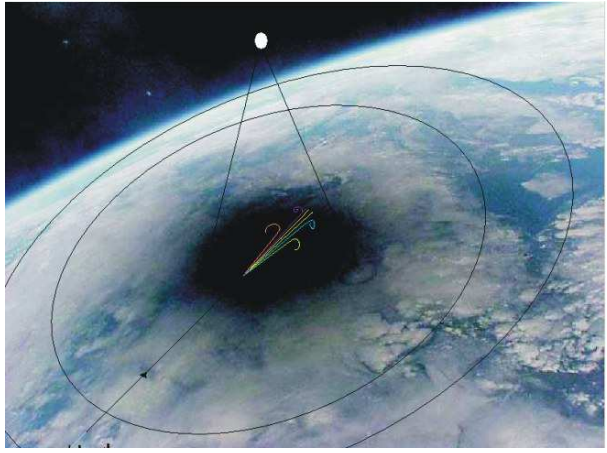}
\caption {A schematic Horizontal High Altitude Shower or similar
Horizontal Tau Air-Shower (HORTAUs) and its open fan-like jets
due to geo-magnetic bending seen from a high quota by EUSO
satellite. The image background is moon eclipse shadow observed
by Mir on Earth. The forked Shower is multi-finger containing a
inner $\gamma$ core and external fork spirals due to $e^+  e^-$
pairs (first opening) and  ${\mu}^+ {\mu}^-$ pairs.}
\label{fig:fig8}
\end{figure}

Any UHE-GZK Tau Air Shower induced event is approximately born
within a wide  ring  (whose radiuses extend between $R \geq 300$
and $R \leq 800$ km from the EUSO Area center). Because of the
wide area and deep $\tau$ penetration \cite{Fargion 2000-2002}
\cite{Fargion 2002b} \cite{Fargion 2002d} the amount of
interacting matter where UHE $\nu$ may lead to $\tau$ is huge
($\geq 2 \cdot 10^5$ $km^3$) ;however only a tiny fraction of
these HORTAUs will beam and Shower within the EUSO Area within
EUSO. We estimate (using also results in \cite{Fargion 2000-2002}
\cite{Fargion 2002b}  \cite{Fargion 2002c} \cite{Fargion 2002d}
\cite{Fargion 2002e} ) an effective Volumes for unitary area (see
schematic Fig. \ref{fig8} below):



\begin{eqnarray}
\frac{V_{eff}}{A_{\oplus}}&=&\int_0^{\frac{\pi}{2}}\frac{(2\,\pi\,\,R_{\oplus}\cos\theta)
\,l_{\tau}\,\sin{\theta}}{4\,\pi\,R^2_{\oplus}}\cdot \nonumber\\
                          &\cdot&e^{-
\frac{2\,R_{\oplus}\,\sin{\theta}}{L_{\nu_{\tau}}}}\,R_{\oplus}\,d\theta\,=
\nonumber\\
&=&\frac{1}{2}\left({\frac{L_{\nu_{\tau}}}{2\,R_{\oplus}}}\right)
^2\,l_{\tau}\int_0^{\frac{2\,R_{\oplus}}{L_{\nu_{\tau}}}}t\cdot
e^{-\,t}d\,t
\end{eqnarray}

Where $V_{eff}$ is the effective volume where Ultra High Energy
neutrino interact while hitting the Earth and lead to escaping
UHE Tau: this volume encompass a wide crown belt, due to the
cross-section of neutrino Earth skimming along a ring of variable
radius $,R_{\oplus}\cos\theta)$ and a corresponding skin crown of
variable depth $l_{\tau}$. $A_{\oplus}$ is the total terrestrial
area, $l_{\tau}$ is the tau interaction lenght, $L_{\nu_{\tau}}$
is the Ultra High Energy Neutrino tau interaction (charged
current) in matter (see in previous Fig. \ref{fig8}). This result
has a very simple approximation, neglecting the small neutrino
absorption \cite{Fargion 2002e}:

\begin{eqnarray}
V_{eff} &=& \frac{V_{eff}}{A_{\oplus}}\cdot A _{Euso} \nonumber \simeq \\
 &\simeq&
 \frac{1}{2}\,A_{Euso}\,\left({\frac{L_{\nu_{\tau}}}{2\,R_{\oplus}}}\right)^2l_{\tau}
\end{eqnarray}

The above geometrical quantities  are  defined in reference
\cite{Fargion 2002b} while $A _{EUSO}$ is the EUSO Area. Assuming
a characteristic EUSO radius of $225$ km and at above energies
one obtains a first rough lower bound:
$$V_{eff}\simeq 3.45 \cdot 10^{3} km^{3}$$
This result is nearly ($\% 40$) less the previous estimate being
here more accurate \cite{Fargion 2002e}.  However the more severe
and realistic suppression factors should be included in present
analytical derivation: first , the exponential decay of air
density at highest (derived in Appendix , see reference
\cite{Fargion 2000-2002}); secondly the Earth Crust opacity at
each integral step, already introduced in previous exact $V_eff$
estimate; the resulting detectable volume from EUSO  becomes:

\begin{eqnarray}
V_{eff}&=&\frac{1}{2}\,A_{Euso}\,\left({1-e^{-
\frac{L_0}{c\,\tau_{\tau}\,\gamma_{\tau}}}}\right)
\,\left({\frac{L_{\nu_{\tau}}}{2\,R_{\oplus}}}\right)^2\cdot
\nonumber\\
          &\cdot&l_{\tau}\left[{1\,-\,e^{-
\frac{2\,R_{\oplus}}{L_{\nu_{\tau}}}}(1\,+\,\frac{2\,R_{\oplus}}{L_{\nu_{\tau}}})
}\right]
\end{eqnarray}


\begin{figure}\centering\includegraphics[width=8cm]{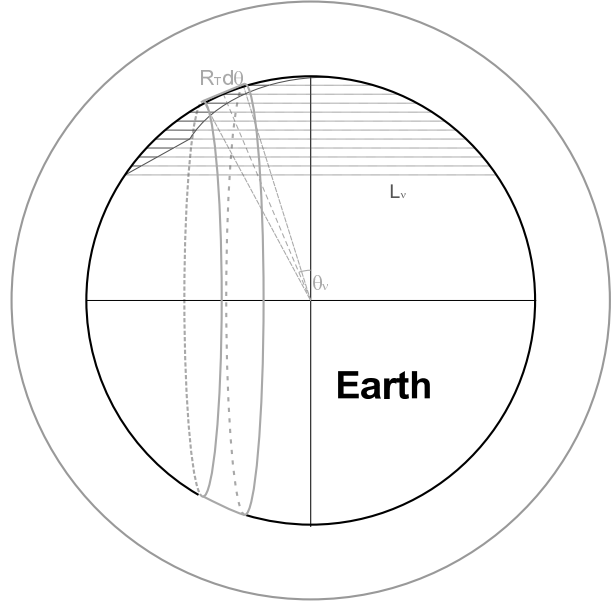}

\caption {A diagram describing the possible incoming neutrino
trajectory and the consequent outcoming tau whose further decay
in flight is the source of HORTAUs within the terrestrial
atmosphere.} \label{fig8}
\end{figure}


The effective Volume, under the assumptions of an incoming
neutrino energy $E_{\nu_{\tau}}= 1.2 E_{\tau}$ and under the
assumption that the outcoming Tau energy leading to a Shower is:
$E_{\tau}= 1.5 E_{Shower}$   corresponds,  in water, at $E_{\tau}=
10^{19} eV$ , to $$V_{eff}=2.662\cdot 10^{3} km^3$$  $$N_{ev} =
20.7$$  at $E_{\tau}= 3\cdot10^{19} eV$, one finds $$N_{ev} =
9.3$$ Finally at $E_{\tau}= 6\cdot10^{19} eV$, $$N_{ev} = 4.5$$
The maximum of the contained events takes place at a Tau energy
$E_{\tau}=1.8\cdot 10^{18} eV$  corresponding to a event number
$$N_{ev} =28.2$$ Therefore from here one would prefer to reduce as much as possible
the EUSO threshold. It should be note that  at $E_{\tau}= 3\cdot
10^{19} eV$ the effective volume reduces to $V_{eff}=8.08\cdot
10^{2} km^3$

The same effective Volume , under the assumptions of an incoming
neutrino energy $E_{\nu_{\tau}}=2 E_{\tau}$ and under the
assumption that the outcoming Tau energy leading to a Shower is
$E_{\tau}= 1.5 E_{Shower}$  corresponds, in water at $E_{\tau}=
10^{19} eV$ , to $V_{eff}=1.83 \cdot 10^{3} km^3$; the
corresponding event number is $N_{ev} = 17$. Assuming an
outcoming Tau at energy $E_{\tau}= 3 \cdot10^{19} eV$,
$V_{eff}=5.5 \cdot 10^{2} km^3$, $N_{ev} = 7.7$.
 The maximum of the volume occurs at a Tau energy $E_{\tau}=2\cdot 10^{17} eV$
and it correspond to a volume $V_{eff}=7.5\cdot 10^{3} km^3$.

\begin{figure}\centering\includegraphics[width=7cm, angle=270]{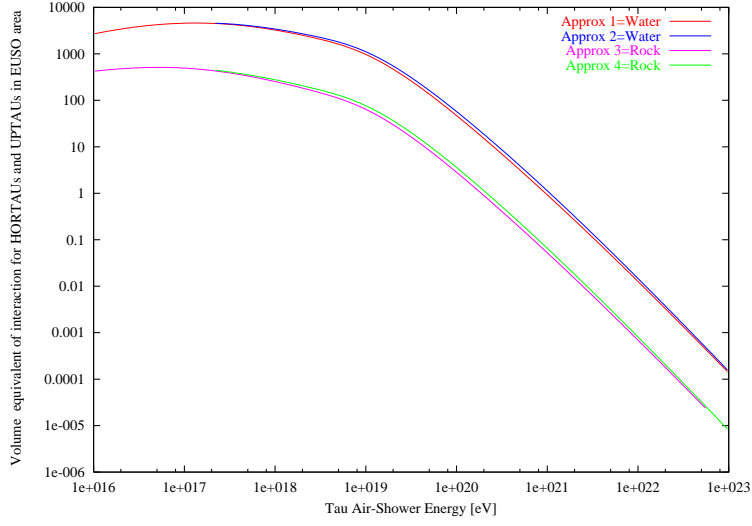}
\caption {Effective Volume of EUSO Event for UPTAUs and HORTAUs .
The effective Volume is corresponding for all Upcoming
Upward-Horizontal Tau Air-Showers, contained within nearly 25 km
altitude. This Volume are under the assumptions of an incoming
neutrino energy $E_{\nu_{\tau}}= 4 E_{\tau}$ and under the
assumption that the outcoming Tau energy leading to a Shower is
$E_{\tau}= 2 E_{Shower}$ }
\end{figure}
The same effective Volume , under the assumptions of an incoming
neutrino energy $E_{\nu_{\tau}}=4 E_{\tau}$ and under the
assumption that the outcoming Tau energy leading to a Shower is
$E_{\tau}= 1.5 E_{Shower}$  corresponds, in water at $E_{\tau}=
10^{19} eV$ , to $V_{eff}=1.13 \cdot 10^{3} km^3$; the
corresponding event number is $N_{ev} = 13.4$. Assuming an
outcoming Tau at energy $E_{\tau}= 3 \cdot10^{19} eV$,
$V_{eff}=3.37 \cdot 10^{2} km^3$, $N_{ev} = 6$.
 The maximum of the volume occurs at a Tau energy $E_{\tau}=2\cdot 10^{17} eV$
and it correspond to a volume $V_{eff}= 4.6\cdot 10^{3} km^3$.
 The maximum of the event number occur at $1.8 10^{18} eV$.

As it is manifest from the above curve the maximal event numbers
takes place at EeV energies. Therefore from here we derived the
primary interest for EUSO to seek lowest threshold (as low as
$10^{19}eV$).

\subsection{UPTAUs volume from HORTAUs formula}

The above expression for the horizontal tau air-shower contains ,
at lowest energies, the UPTAUs case. Indeed it is possible to see
that the same above  effective volume at lowest energies simplify
and  reduces to:

\begin{eqnarray}
V_{eff}&=&\frac{1}{2}\,A_{Euso}\,\left({1-e^{-
\frac{L_0}{c\,\tau_{\tau}\,\gamma_{\tau}}}}\right) \,l_{\tau}
\end{eqnarray}

Because one observes the Earth only from one side   the Area
factor in eq. $1$ should be $A_{\oplus} = {2\,\pi\,R^2_{\oplus}}$
and therefore the half in above formula may be dropped and the
final result is just the common expression $V_{eff} =
A_{Euso}\,l_{\tau}$.

\section{Event Rate for GZK neutrinos at EUSO}

The above effective volume should be considered for any given
neutrino flux to estimate the outcoming EUSO event number. Here
we derive first the analytical formula. These general expression
will be plot assuming a minima GZK neutrino flux $ \phi_{\nu}$
just comparable to observed UHECR one $ \phi_{\nu}\simeq
\phi_{UHECR} \simeq 3\cdot 10^{-18} cm^{-2} s^{-1} sr{-1}$ at the
same energy ($E_{\nu}= E_{UHECR}\simeq 10^{19} eV$). This
assumption may changed at will (model dependent) but the event
number will scale linearly accordingly to the model. From here we
may estimate the event rate in EUSO by a simple extension:

\begin{eqnarray}
N_{eventi}\,=\,\Phi_{\nu}\,4\,\pi\,\eta_{Euso}\Delta
\,t\,\left({\frac{V_{eff}}{L_{\nu}}}\right)
\end{eqnarray}

Where $\eta_{Euso}$ is the  duty cycle fraction of EUSO,
$\eta_{Euso} \simeq 10\%$, $\Delta \,t\ \simeq 3$  $years$ and
$L_{\nu}$ has been defined in Fig. \ref{fig4}.


\begin{eqnarray}
N_{eventi}&=&\Phi_{\nu_\tau}\,A_{Euso}\,(\frac{1}{2}\,4\pi\,\eta_{Euso})\,{\Delta{t}}\cdot \nonumber\\
          &\cdot&\left({1-e^{-\frac{L_0}{c\,\tau_{\tau}\,\gamma_{\tau}}}}\right)\left({\frac{l_{\tau}}{L_{\nu_{\tau}}}}\right)\left({\frac{L_{\nu_{\tau}}}{2\,R_{\oplus}}}\right)^2\cdot \nonumber\\
          &\cdot&\left[{1\,-\,e^{-\frac{2\,R_{\oplus}}{L_{\nu_{\tau}}}}(1\,+\,\frac{2\,R_{\oplus}}{L_{\nu_{\tau}}})}\right]
\end{eqnarray}

It should be remind that all these event number curves for EUSO
are already suppressed by a factor $0.1$ due to minimal EUSO duty
cycle.






\begin{figure}\centering\includegraphics[width=7cm, angle=270]{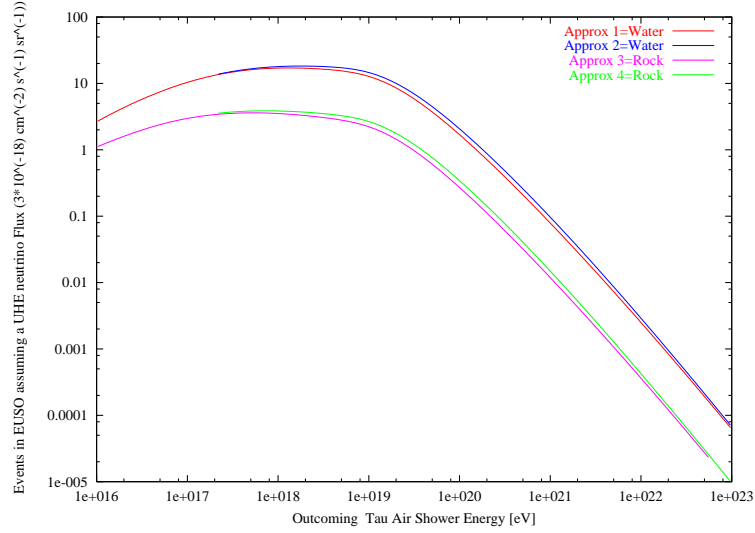}
\caption {Number of EUSO Event for HORTAUs in 3 years record. The
incoming neutrino has an energy $4$ larger the outcoming Tau;
this born Tau has an energy $2$ times the final Tau Air-Shower
end energy .}
\end{figure}

\begin{figure}\centering\includegraphics[width=12cm]{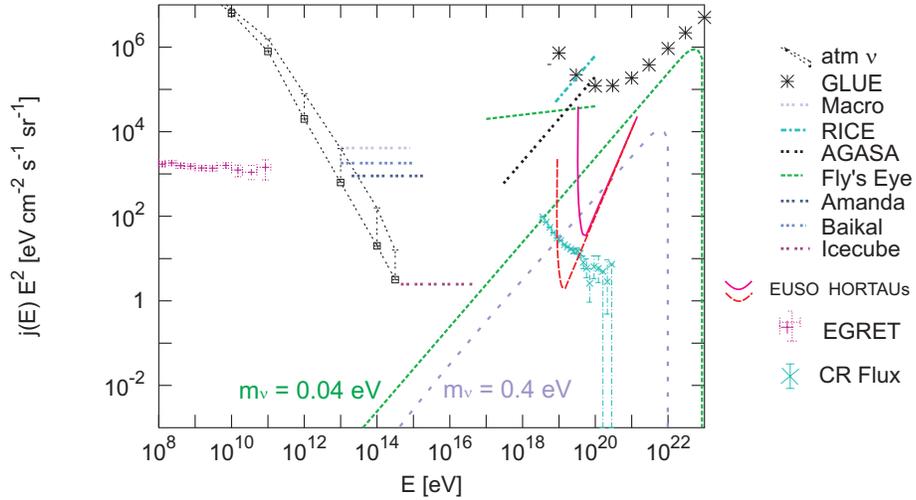}
\caption {EUSO thresholds for Horizontal Tau Air-Shower HORTAUs
(or Earth Skimming Showers) over few $\gamma$, $\nu$ and Cosmic
Rays (C.R.) Fluence and bounds. Continuous and dashed curve for
HORTAUs are drawn assuming respectively an EUSO threshold at
$3\cdot 10^{19}$eV and at lower $10^{19}$eV values. Because the
bounded $\tau$ flight distance (due to the contained terrestrial
atmosphere height) the main signal is better observable at $1.1
\cdot 10^{19}$eV than higher energies. The Fluence threshold for
EUSO has been estimated for a three year experiment lifetime.
Z-Shower or Z-Burst expected spectra in light neutrino mass values
($m_{\nu} = 0.04, 0.4$ eV) are shown. } \label{fig:fig9}
\end{figure}


\begin{figure}\centering\includegraphics[width=12cm]{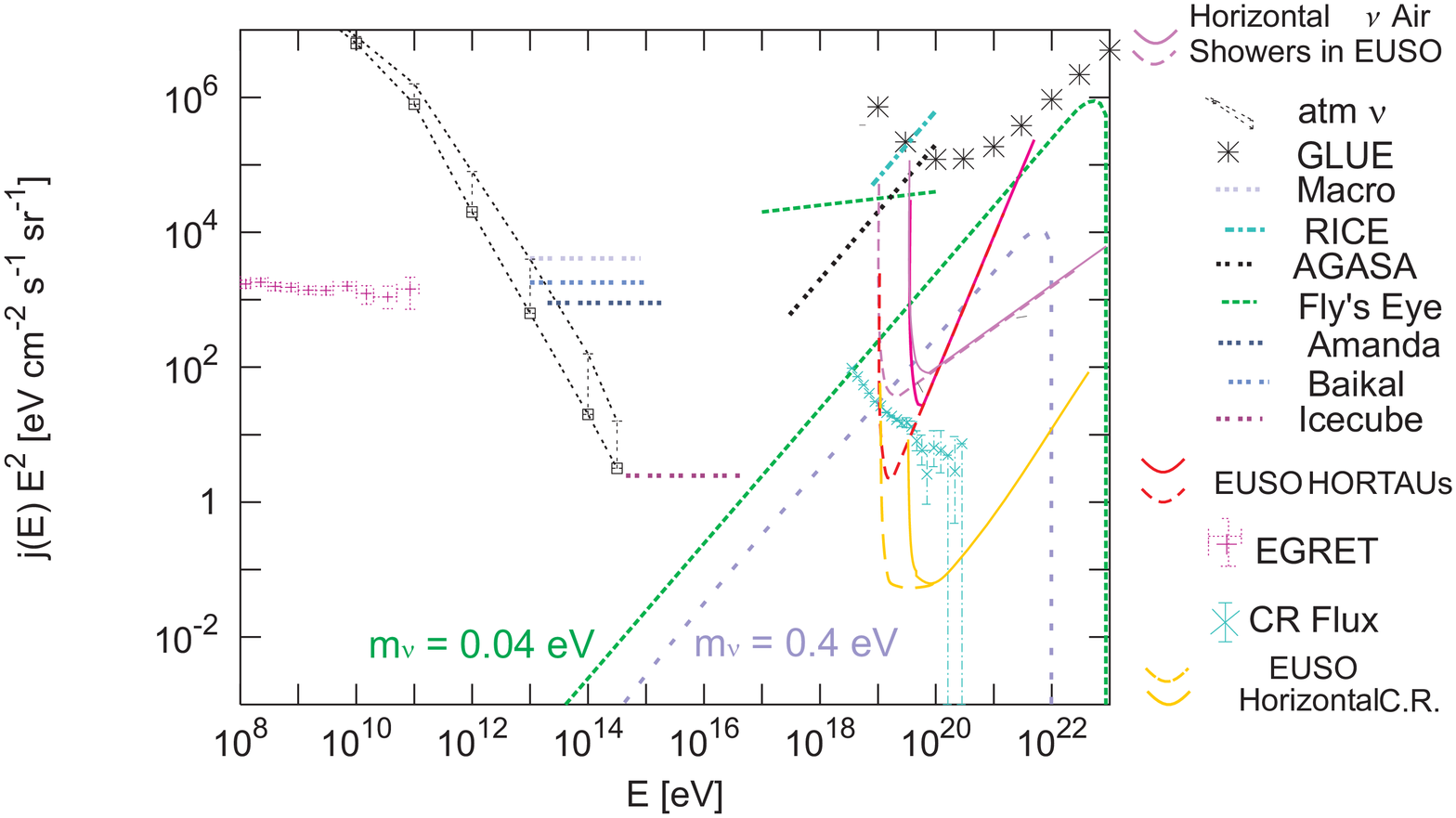}
\caption {EUSO thresholds for Horizontal Tau Air-Shower shower,
HORTAUs (or Earth Skimming Showers) over all other $\gamma$, $\nu$
and Cosmic Rays (C.R.) Fluence and bounds. The Fluence threshold
for EUSO has been estimated for a three year experiment lifetime.
Competitive experiment are also shown as well as the Z-Shower
expected spectra in light neutrino mass values ($m_{\nu} = 0.04,
0.4$ eV). As above dash curves for both HORTAUs and Horizontal
Cosmic Rays are drawn assuming an EUSO threshold at $10^{19}$eV. }
\end{figure}









However the Air-Shower induced neutrino may reflect all three
light neutrino flavours, while HORTAUs are made only by
$\nu_{\tau}$,$\bar{\nu_{\tau}}$ flavour. Nevertheless the
dominant role of HORTAUs overcome (by a factor $\geq 15$) all
other Horizontal EUSO neutrino event: their  expected event rate
are, at $\Phi_{\nu}\geq 3 \cdot 10^{3}$ eV $cm^{-2} s^{-1}$
neutrino fluence (as in Z-Shower model in Figure $13-18-19$), a
few hundred event a year and they may already be comparable or
even may exceed the expected Horizontal CR rate. Dash curves for
both HORTAUs and Horizontal Cosmic Rays are drawn assuming an EUSO
threshold at $10^{19}$eV. Because the bounded $\tau$ flight
distance (due to the contained terrestrial atmosphere height) the
main signal is  better observable at $1.1 \cdot 10^{19}$eV than
higher energies as emphasized  in Fig.$13-18$ at different
threshold curves.

However the Air-Shower induced neutrino may reflect all three
light neutrino flavours, while HORTAUs are made only by
$\nu_{\tau}$,$\bar{\nu_{\tau}}$ flavour. Nevertheless the
dominant role of HORTAUs overcome (by a factor $\simeq 6$) all
other Horizontal EUSO neutrino events: their  expected event rates
are, at $\Phi_{\nu}\geq 3 \cdot 10^{3}$ eV $cm^{-2} s^{-1}
sr^{-1}$ neutrino fluence (as in Z-Shower model ), a few hundred
event a year and they may already be comparable or even may
exceed the expected Horizontal CR rate. Dash curves for both
HORTAUs and Horizontal Cosmic Rays are drawn assuming an EUSO
threshold at $10^{19}$eV. Because the bounded $\tau$ flight
distance (due to the contained terrestrial atmosphere height) the
main signal is better observable at $1.1 \cdot 10^{19}$eV than
higher energies as emphasized   at different threshold curves.

\section{Partially, Fully contained and Crossing events}
  The HORTAUs are very long showers. Their lenght may exceed
  two hundred kilometers. This trace may be larger than the EUSO
  radius of Field of View. Therefore there may be both contained
  and partially contained events. There may be also crossing
  HORTAUs at the edges of EUSO  disk area. However most
  of the events will be partially contained, either just on their
  birth or at their end, equally balanced in number. Because of the HORTAU Jet forked shower,
  its up-going direction, its fan like structure, these partially
  contained shower will be the manifest and mostly useful and
  clear event. The area of their origination, four times larger than
  EUSO field of view, will be mostly
  outside the same EUSO area. Their total number count will double the
  event rate $N_{ev}$ (and the corresponding $V_{eff}$) of
  HORTAUs. The additional crossing event will make additional events (a small fraction) of the effective
  volume of HORTAUs at $10^{19} eV$ the most rich neutrino signal few times larger the Air induced events.
  The same doubling will apply only to UHECR horizontal shower
  while the downward Ultra High Energy Neutrino will not share
  this phenomena (out of those $\simeq 6\%$ of the Horizontal Air Neutrino
  Shower).

Highest Energy Neutrino signals may be well observable by next
generation satellite as EUSO: the main revolutionary source of
such neutrino traces are UPTAUs (Upward Tau blazing the telescope
born in Earth Crust) and mainly HORTAUs (Horizontal Tau
Air-Showers originated by an Earth-Skimming UHE $\nu_{\tau}$).
There are still confusing questions about the Air or the Earth
Crust role as the ruling neutrino calorimeter. The role of
downward Air Showers Induced by Neutrinos is apparently related
to its impressive volume($V_{Air}\simeq$$1500 km^3$). However
these down ward Showers will be  observable in a limited solid
angle (inclined) ($V_{Air}\leq $$500 km^3$) at lowest quota.
Because of the air opacity their fluorescent lights will be mostly
suppressed. Their signal will be naturally drown in UHECR ruling
noise. There is , moreover for these down-ward neutrino, a much
better competitive calorimeter in Auger; indeed this experiment
(and maybe its twin in North America) might well observe the
downward Neutrino Air showering and the Auger Area ($3000 km^3$)
with no duty cycle cut-off, and within its full solid angle view,
it may well compete and even it will exceed the effective volume
of Air in Field of View of EUSO.  On the other side lower
energetic UPTAUs (at PeV energies because the Earth neutrino
opacity at higher energies) will be rarely blazing to EUSO
because of the short $\tau$ boosted lenght, at Pevs
energies\cite{Fargion 2000-2002}; most of these UPTAUs events
occur on Earth crust (by a factor of four-five) over other events
taking place in air. UPTAUs will be detected at the boundary of
EUSO disk as a thin stretched small multi-pixel line, whose
orientation is strongly "polarized" orthogonal to the local
geo-magnetic field. The EUSO sensibility (effective volume
($V_{eff}$$\sim 0.1 km^3$) will be nearly just an order of
magnitude above present AMANDA-Baikal bounds.
 However the most relevant  Horizontal
Tau Air-Shower, HORTAUs at GZK energies will be better searched
and  revealed at best at lowest EUSO thresholds. They are
originated along huge Volumes around the EUSO Area. Their
horizontal skimming secondary $\tau$ decay occur far away $\geq
550$ km, at high altitude ($\geq 20-40$ km) and they will give
clear signals very distinguished from any downward horizontal
UHECR. There are none of such upward UHECR event expected.
Therefore there is not noise background.  HORTAUs are grown by UHE
neutrino interactions inside huge volumes ($V_{eff}$$\geq 2300
km^3$)  for incoming neutrino energy $E_{\nu_{\tau}}$ $\simeq
10^{19}$ eV.
\begin{figure}
\centering
\includegraphics[width=12cm]{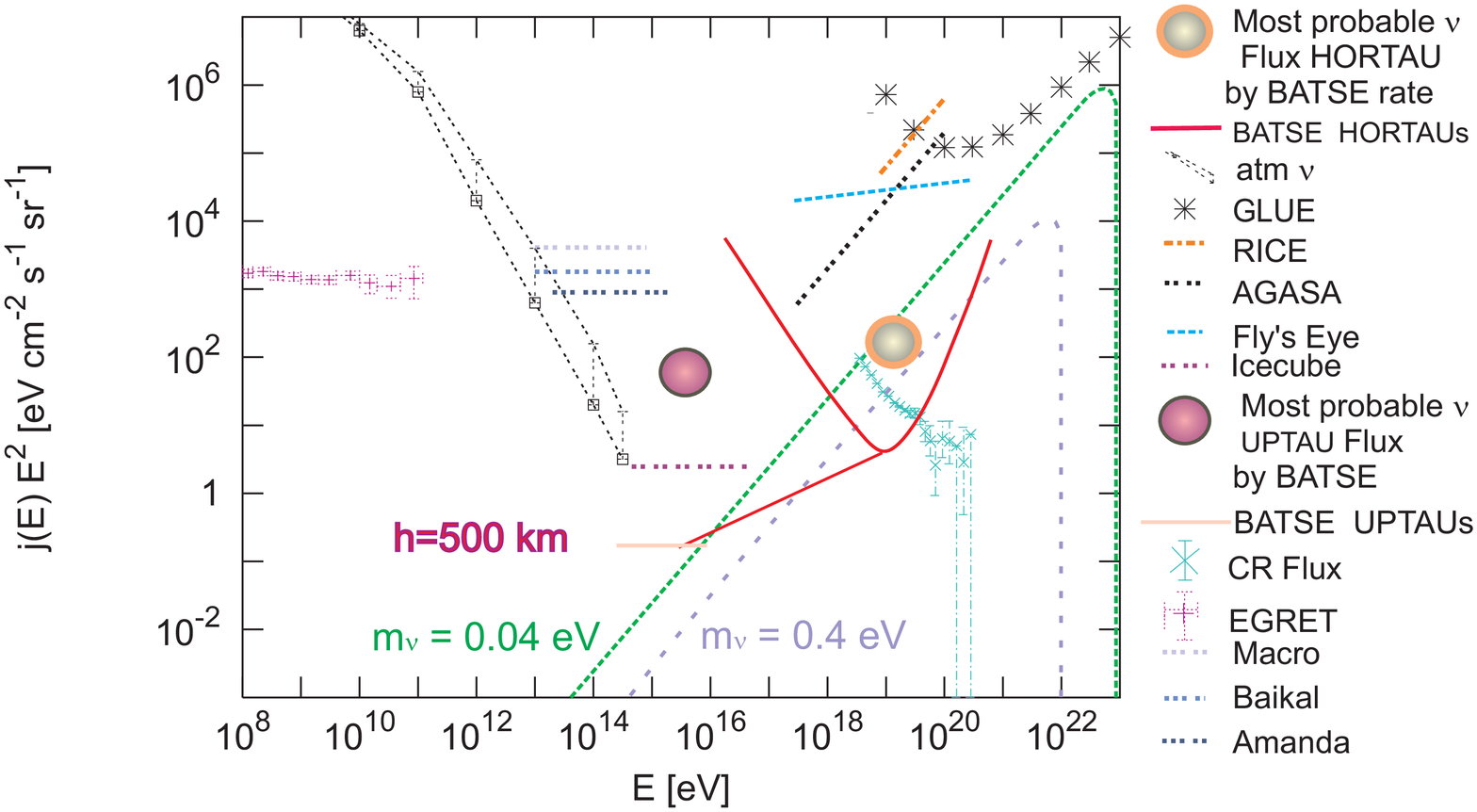}
\caption {Neutrino Flux derived by  BATSE Terrestrial Gamma
Flashes assuming them as $\gamma$ secondaries of upward Tau
Air-showers. These fluxes are estimated using the data from
Terrestrial Gamma Flash (1991-2000) normalized during  their most
active trigger and TGF hard activities. The UPTAUs and HORTAUs
rate are normalized assuming that the events at geo-center angle
above $50^o$ might be of HORTAU nature. } \label{fig:fig33}
\end{figure}

 Even for the most conservative scenario where a minimal GZK-$\nu$ fluence must
take place (at least at:$E_{\nu}\,\Phi_{\nu} \simeq 30 eV cm^{-2}
s^{-1}sr^{-1}$, just comparable to well observed Cosmic Ray
fluence), a dozen or more of such UHE astrophysical neutrino must
be observed  during three year of EUSO data recording. These
HORTAUs will be not be observable by other competitive experiment
as AUGER. Therefore to improve the HORTAU visibility in EUSO one
must,  : a) Improve the fast pattern recognition of Horizontal
Shower Tracks with their few distant dots with forking signature.
 b) Enlarge the Telescope Radius to embrace also lower $10^{19}$ eV
 energy thresholds where UHE HORTAU neutrino signals are enhanced.
 c) Consider a detection  at  angular $\Delta\theta$ and at height $\Delta h$ level within an
accuracy $\Delta\theta \leq 0.2^o$, $\Delta h \leq 2$ km.\\ Even
all  the above results have been derived carefully  following
\cite{Fargion 2002b} \cite{Fargion 2002c} \cite{Fargion 2002e} in
a minimal realistic framework they may be used within $20\%$
nominal value due to the present uncertain in  EUSO detection
capabilities.

 \section{Conclusions}
 UHECR and Neutrino Astronomy face a new birth.
The Neutrino Astronomy may be widely discovered by Upward and
Horizontal $\tau$ Air-Showers. The Tau neutrinos , born
abundantly by flavour mixing will probe such Astronomy above PeVs
up to EeVs energies, where astrophysics rule over atmospheric
neutrino noise. The same UHE $\overline{\nu_{e}}$  at $E_{\nu_{e}}
= \frac{{M_W}^2}{2 \cdot m_e} \simeq 6.3 PeV $  must be a peculiar
neutrino astronomy born beyond Mountain Chains \cite{Fargion et
all 1999},\cite{Fargion 2000-2002} with its distinctive signature.
Past detectors as GRO BATSE experiment might already found some
direct signals of such rare UPTAUs or HORTAUs; indeed their
observed Terrestrial Gamma Flash event rate translated into a
neutrino induced  upward shower (see Fig.$19$) leads to a most
probable flux both at PeVs energies  just at a level comparable
to most recent AMANDA threshold sensitivity: for horizontal TGF
events at $10^{19}$ eV windows, the signals fit the Z-Burst model
needed fluence (for neutrino  at $0.04-0.4$ eV masses).
 Future EUSO telescope detector, if little enlarged will easily probe even the smallest, but necessary
Neutrino GZK fluxes with clear sensitivity (see Fig $17-18$). We
therefore expect that a serial of experiment will soon turn
toward this last and neglected, but most promising Highest Energy
Neutrino Tau Astronomy searching for GZK or Z-Showers neutrino
signatures.
\subsection{Acknowledgment}
The author wish to thank Prof. M. Baldo Ceolin for the very
exciting meeting in Venice and the discussions as well as
G.Salvini for kind support and suggestions.

%
\bibliography{xbib}
\end{document}